\documentstyle[12pt]{article}
\makeatletter
\@addtoreset{equation}{section}
\makeatother

\begin{document}
\begin{center}
{\Large \bf
Indeterministic Quantum Gravity and Cosmology} \\[0.5cm]
{\large\bf X. Probability-Theoretic Aspect:\\
A Hidden Selector for Quantum Jumps,\\
or How the Universe Plays the Game of Chance}\\[1.5cm]
{\bf Vladimir S.~MASHKEVICH}\footnote {E-mail:
mash@gluk.apc.org}  \\[1.4cm]
{\it Institute of Physics, National academy
of sciences of Ukraine \\
252028 Kiev, Ukraine} \\[1.4cm]
\vskip 1cm

{\large \bf Abstract}
\end{center}

This paper is a sequel to the series of papers [1-9]. The
problem of the meaning of objective a priori probability for
individual random trials without repetition is considered.
A sequence of such trials, namely quantum jumps, is
realized in indeterministic dynamics of the universe. A
hidden selector for the quantum jumps is constructed.

\newpage

\hspace*{9 cm}
\begin{minipage} [b] {9 cm}
It is as natural as natural selection...
\end{minipage}
\begin{flushright}
William Golding \vspace*{0.8 cm}
\end{flushright}

\begin{flushleft}
\hspace*{0.5 cm} {\Large \bf Introduction}
\end{flushleft}

Indeterminism implies objective probability, so that
indeterministic quantum gravity and cosmology (IQGC)---
the theory being developed in this series of papers---
relies on probability theory. In this regard, IQGC is
notably specific. Generally the experimental verification
of probability-theoretic results is considered in the
context of identical conditions repeated many times. In
IQGC (and not only in it), the repetitions are, in
general, unavailable. Here we are faced with the problem
of the meaning of objective a priori probability for
individual situations. Thus the probability-theoretic
aspect of IQGC should be specially studied.

The most solid way for handling the problem of individual
situations is to get around it, namely, to reduce the
choice in a sequence of the individual situations to the
employment of the results of the sequence of one and the
same random trial. In the case of IQGC this is done as
follows.

The sequence of individual situations is that of quantum jumps
numbered by $i\in Z=\left\{ 0,\pm 1,\pm 2,... \right\}$. A
result of the $i$-th jump is one of the two states [5]:
$\omega_{i}^{\pm}$ with energies $\epsilon_{i}^{\pm},
\;\epsilon_{i}^{+}>\epsilon_{i}^{-}$, and a priori probabilities
$w_{i}^{\pm}$.

We take the interval $[ 0,1 ]$ and divide it into two
parts: $\Delta_{i}^{-}=[0,w_{i}^{-}],
\;\Delta_{i}^{+}=(w_{i}^{-},1]$. The random trial is the choice of
a point from $\left[ 0,1 \right]$, the distribution being uniform.
Denote the results $\omega_{i}^{\pm}$ of the $i$-th jump by
$e_{i}=\pm 1$. Let the result of the $i$-th trial be $x_{i}$. Then
$e_{i}=\chi_{\Delta_{i}^{+}}(x_{i})-\chi_{\Delta_{i}^{-}}(x_{i})$
where $\chi$ is the characteristic function.

Here we have the Bernoulli trials [10], the sequence $S=
\left\{ x_{i}:i\in Z \right\}$ being its choice function, or
selector. The latter is a hidden selector for quantum jumps in
IQGC.

Now the universe $U$ as a dynamical system is $U=(M,S,\omega)$,
where $M$ is a fixed spacetime manifold [6] and $\omega$ is a
dynamical state: $\omega=(g,\dot g, \Psi)$, $g$ is a metric
and $\Psi$ is a state vector of matter.

\section{Indeterministic dynamics of the universe as a
choice function of a stochastic process}

In IQGC we have a denumerable set of quantum jumps,
\begin{equation}
J=\left\{ e_{i}:i\in Z \right\}, \quad Z=0,\pm 1, \pm 2,...,
\label{1.1}
\end{equation}
where $e_{i}$ stands for the $i$-th event, i.e., the jump.
A result of the $i$-th jump is one of the two admissible states
$\omega_{i}^{\pm}$ of matter [5] with energies
\begin{equation}
\epsilon_{i}^{\pm},\qquad \epsilon_{i}^{+}>\epsilon_{i}^{-}
\label{1.2}
\end{equation}
and a priori probabilities
\begin{equation}
w_{i}^{\pm},\qquad w_{i}^{+}+w_{i}^{-}=1.
\label{1.3}
\end{equation}
It is convenient to put
\begin{equation}
e_{i}=\left\{
       \begin {array}{rcl}
	+1\quad {\rm for}\quad \omega_{i}^{+}\\
	-1\quad {\rm for}\quad \omega_{i}^{-}.\\
       \end{array}
      \right.
\label{1.4}
\end{equation}
Then we have
\begin{equation}
J=\left\{ e_{i}\in\left\{ +1,-1 \right\}:i\in Z \right\}.
\label{1.5}
\end{equation}

We may regard $J$ as a choice function of a stochastic process.
In fact, $J$ is the choice function: it is the only one
realized, there are no repetitions for the process. Thus we
are faced with the first problem---that of the meaning of
objective a priori probabilities (\ref{1.3}) in the case
when only one choice function is realizable.

\section{Probabilistic interpretation of a family of
random trials without repetition}

A natural general solution to the first problem was proposed
in [11].

Let
\begin{equation}
\left\{ (X_{\alpha},A_{\alpha},P_{\alpha}):
\alpha=1,2,...,N \right\}
\label{2.1}
\end{equation}
be a family of probability spaces. Then
\begin{equation}
f(x_{1},x_{2},...,x_{N})=\frac{1}{N}\sum_{\alpha}^{1,N}
f_{\alpha}(x_{\alpha}),\qquad x_{\alpha}\in X_{\alpha},
\label{2.2}
\end{equation}
is a random variable. A single measurement of $f$ is performed,
i.e., a single value for each $x_{\alpha}\quad (\alpha=
1,2,...,N)$ is obtained. Then for $N\gg 1$, the value of $f$
obtained experimentally is close to the mean value:
\begin{equation}
f\approx \langle f \rangle=\frac{1}{N}\sum_{\alpha}^{1,N}
\langle f_{\alpha}\rangle.
\label{2.3}
\end{equation}
Eq.(\ref{2.3}) defines the meaning of a priori probability for
individual situations.

The usual approach corresponds to the case where
\begin{equation}
(X_{\alpha},A_{\alpha},P_{\alpha})=(X,A,P),\quad
\alpha=1,2,...,N,\qquad f_{1}=f_{2}=...=f_{N}.
\label{2.4}
\end{equation}

\section{The problem of the selection of a choice function}

Although the approach given in the previous section
elucidates the meaning of objective a priori probability
for individual situations, we remain faced with the second
problem: Is there a mechanism of the selection of an actual
choice function $J$ (\ref{1.5}), which would grant that
eq.(\ref{2.3}) be fulfilled?

\section{The Bernoulli trials and a hidden selector for
quantum jumps}

The most solid solution to the second problem consists in
the employment of the results of the sequence of one and the
same random trial. In other words, we employ the Bernoulli
trials.

The random trial is the choice of a point $x$ from the interval
$\left[ 0,1 \right]$, the distribution being uniform:
\begin{equation}
{\rm for}\quad \left[ a,b \right]\subset \left[ 0,1 \right]
\qquad P(\left[ a,b \right])=b-a.
\label{4.1}
\end{equation}
Let
\begin{equation}
S=\left\{ x_{i}:i\in Z \right\}
\label{4.2}
\end{equation}
be a given choice function of the Bernoulli trials. For each
$i$ we divide the interval $\left[ 0,1 \right]$ into two
parts:
\begin{equation}
\Delta_{i}^{-}=[0,w_{i}^{-}],\qquad \Delta_{i}^{+}=
(w_{i}^{-},1].
\label{4.3}
\end{equation}
Now we put
\begin{equation}
e_{i}=\chi_{\Delta_{i}^{+}}(x_{i})-\chi_{\Delta_{i}^{-}}
(x_{i}),
\label{4.4}
\end{equation}
where $\chi$ is the characteristic function, so that the
result of the $i$-th quantum jump is determined by the
following prescription:
\begin{equation}
\omega_{i}=\left\{
	    \begin{array}{rcl}
	     \omega_{i}^{+}\quad {\rm for}\quad x_{i}
	     \in\Delta_{i}^{+}\\
	     \omega_{i}^{-}\quad {\rm for}\quad x_{i}
	     \in\Delta_{i}^{-}.\\
	    \end{array}
	   \right.
\label{4.5}
\end{equation}

Thus the choice function $S$ (\ref{4.2}) of the Bernoulli
trials serves as a hidden selector for quantum jumps.

\section{The universe as a dynamical system}

The description of the universe $U$ as a dynamical system
may be given as follows. $U$ is the triple,
\begin{equation}
U=(M,S,\omega);
\label{5.1}
\end{equation}
$M$ is a fixed spacetime manifold described in [6]; $S$ is
a hidden selector; $\omega$ is a dynamical state,
\begin{equation}
\omega=(g,\dot g,\Psi)
\label{5.2}
\end{equation}
where $g$ is a metric and $\Psi$ is a state vector of matter;
the metric is of the form [5,6]
\begin{equation}
g=dt\otimes dt-h_{t},
\label{5.3}
\end{equation}
so that
\begin{equation}
\omega=(h,\dot h,\Psi).
\label{5.4}
\end{equation}

\section*{Acknowledgment}

I would like to thank Stefan V. Mashkevich for helpful
discussions.

\end{document}